\documentclass[dvips]{acta}
\usepackage{supertabular,lscape,epsfig}
\usepackage{amssymb}
\usepackage{amsmath}
\usepackage[T1]{fontenc}

\SetPages{0}{0}

\SetVol{76}{2026}

\usepackage{lmodern}

\newcommand{\NS}{N_{\rm s}}
\newcommand{\NC}{N_{\rm in}}
\newcommand{\NT}{N_{\rm str}}
\newcommand{\WS}{W_{\rm s}}
\newcommand{\WC}{W_{\rm c}}
\newcommand{\WT}{W_{\rm t}}
\newcommand{\PHS}{\Phi_{\rm surf}}
\newcommand{\PHB}{\Phi_{\rm bot}}
\newcommand{\RES}{\Re(\sigma)}
\newcommand{\IMS}{\Im(\sigma)}
\newcommand{\mFig}[1]{Fig.~\ref{fig:#1}}
\newcommand{\mEq}[1]{Eq.~(\ref{eq:#1})}
\newcommand{\mTab}[1]{Table~\ref{tab:#1}}

\DeclareMathOperator{\sgn}{sgn}

\begin{document}

\begin{Titlepage}

\Title{A quadratic balance relation for radial nonadiabatic pulsations}

\Author{Zalewski, J.,}
{Independent researcher \\
e-mail: jan.zalewski.a2@gmail.com}

\Received{July 28, 2026}
\end{Titlepage}

\Abstract{
	Using a sesquilinear pressure - rate of volume change kernel and linear pulsation equations, we formulate a balance relation combining work integral, surface terms, norm integral and the real part of pulsation frequency (for $\exp({\omega t})$ time dependence) in terms of linear radial pulsation variables. The resulting relation is exact within the adopted radial nonadiabatic formulation. We apply the quadratic balance relation to radial pulsations in AGB envelopes and use it to check the accuracy of computation and to obtain information about mode excitation and damping. We also compare the balance relation with the classical formulation based on dissipation and kinetic energy integrals and discuss the meaning of the components made explicit by the present decomposition that appear in our balance equation and their relevance for ordinary and strange modes in AGB envelopes.
}

{stars: AGB and post-AGB, stars: oscillations, stars: interiors, methods: numerical}

\section{Introduction}
Work integrals and kinetic energy integrals have long been used as diagnostics of excitation and damping in stellar pulsation theory (Cox (1980), Unno et al. (1989)). The work integral provides interpretation of pulsational stability in terms of driving and damping regions and is widely used in the interpretation of pulsational instabilities of various types of stars (see Dziembowski, 1994, Pamyatnykh, 1999). In conjunction with kinetic energy integral it leads to a relation from which an estimate of the mode excitation rate may be obtained. This derivation, however, is approximately valid only for weakly nonadiabatic pulsations and cannot be applied without qualification to highly nonadiabatic pulsations in supergiant envelopes.

Alternative formulations have been developed for cases in which the conventional cycle average is inadequate. Buchler and Regev (1982) used a multiple-time formalism to define averaged work for slowly evolving pulsations. Glatzel (1994) derived a mechanical energy balance from the Lagrangian continuity and momentum equations and replaced the cycle average by an ensemble average over the arbitrary initial phase of a complex eigenfunction. His formulation remains applicable when the growth or damping time is comparable to the pulsation period.

In the present paper we recast the mechanical balance in terms of the radial nonadiabatic linear system of Dziembowski (1977). Instead of deriving the formulation from ensemble of real eigensolutions we postulate the sesquilinear pressure - volume rate kernel as suggested by $P\,dV$. The real part of this kernel is used as the work-like quadratic amplitude form. When it is applied to the radial nonadiabatic first-order system an exact quadratic balance relation is obtained. The relation involves work integral, surface terms, and a generalized norm integral related through the real part of complex eigenfrequency. It is formally analogous to the classical work - energy relation for pulsations, but contains additional terms whose origin and role will be discussed below.

The purpose of the paper is twofold. First, to formulate the balance relation using Dziembowski (1977) pulsation equations and compare its terms with the classical work and kinetic energy integrals. Second, apply the resulting diagnostics to ordinary, strange and thermal modes in AGB-envelope models. We analyze the terms appearing in the relation and examine their relevance for the balance for radial ordinary and strange modes.

\section{Pulsation equations and required identities}
Since in what follows we will focus primarily on radial pulsations in the envelopes of AGB/post-AGB stars, we adopt the variables introduced by Dziembowski (1977). For radial oscillations, the perturbation of the gravitational field can be eliminated exactly in terms of $d$, so the pulsation problem closes in terms of the following four variables
\[
d=\frac{\delta r}{r},
\qquad
p=\frac{\Delta P}{P},
\qquad
s=\frac{\Delta S}{c_P},
\qquad
f=\frac{\Delta L}{L}.
\]

The independent variable is

\[
x=\ln(r/R_\odot),
\]
and the eigenfunctions are normalized by setting $d(x_{\rm s})=1$.

With this the linearized nonadiabatic pulsation equations may be written as

\begin{equation}
y' = M(x,\sigma)\,y
\label{eq:basicEq}
\end{equation}
where $\sigma=\omega/\sqrt{4 \pi G \langle\rho\rangle}$ is the non-dimensional, complex frequency, and the time dependence of perturbations is $\exp{(\omega\, t)}$. The definitions of the coefficients $\texttt{A}_{i=1,13}$ used to compute the pulsation matrix $M$ may be found in Dziembowski (1977). For radial pulsations the relation $w_1=-A_5d$ eliminates the gravitational variables and removes the corresponding $A_5$ terms from $M$.

We also introduce derived perturbation quantities for volume $q$ and temperature $t$ perturbations, by expressing them in terms of $p$ and $s$ and $\texttt{A}$ as
\[
\begin{aligned}
q&=\frac{\Delta V}{V} =-\frac{\Delta \rho}{\rho} =-\texttt{A}_4\, p-\texttt{A}_7\,s, \\
t&=\frac{\Delta T}{T} =\nabla_{ad}\,p+s=\texttt{A}_8\texttt{A}_{10}\,p+s
\end{aligned}
\]
with $\texttt{A}_4=1/\Gamma_1$, $\texttt{A}_7=\left(\frac{\partial \ln{\rho}}{\partial \ln{T}}\right)_P$, $\texttt{A}_8=\nabla$ and $\texttt{A}_{10}=\nabla_{ad}/\nabla$. 

With these definitions in place we may now proceed to the derivation of the quadratic balance relations associated with the nonadiabatic pulsation equations.

\section{Derivation of the quadratic balance relation}

In this section we derive a global balance relation associated with the nonadiabatic pulsation equations given by \mEq{basicEq}. At this point we do not make any assumptions about the nature of the pulsations, such as near-adiabatic behavior, as is done in the classical derivation. Instead, the relevant quadratic forms will be derived directly from the equations of pulsation.

\subsection{Derivation of the work integral}

Motivated by the work done on a fluid element, $-P\,dV$, we introduce a second-order pairing between the pressure perturbation and the rate of change of the volume perturbation. In terms of the relative pressure and volume perturbations this complex pressure and volume-rate pairing may be written as
\[
{\cal P}=-p\,\overline{\dot q}.
\]

Using $\dot q=\frac{dq}{dt}=\omega q=\sqrt{4 \pi G\langle \rho \rangle}\sigma q$ we introduce the nondimensional complex work kernel
\begin{equation}
{\cal K}_{\rm s} = -p\,\overline{\sigma q}
\label{eq:Kay}
\end{equation}
and include the frequency scaling factor in the $C(x)$ defined below. 

We associate with this kernel the following real amplitude-level work contribution
\[
\frac{1}{2}\Re({\cal K}_{\rm s}).
\]
The real part is retained because the balance relation sought here concerns the growth rate, $\RES$, in analogy with the classical relation between the dissipation and kinetic energy integrals.

Using this and for a radial shell $dV_0=4\pi r^3\, dx$ the luminosity-scaled work integrand is
\[
\frac{d\WS}{dx}=\frac{1}{2}4\pi r^3 P\frac{\sqrt{4 \pi G \langle \rho \rangle}}{L} \Re({\cal K}_{\rm s}).
\]

Defining
\[
C(x)=\frac{1}{2}4 \pi r^3 P \frac{\sqrt{4\pi G\langle \rho\rangle}}{L}
\]
we obtain
\[
\frac{d\WS}{dx}=C(x)\Re({\cal K}_{\rm s}).
\]

In the notation used in the pulsation equations the $C$ can be expressed as
\[
C(x) = \frac{1}{2}\texttt{A}_{13}\frac{\texttt{A}_8\texttt{A}_{10}}{-\texttt{A}_7}.
\]

Hence the scaled work integral $\WS$ is given by

\[
\WS = \int^{x_{\rm s}}_{x_b} C(x) \Re({\cal K}_{\rm s}) dx.
\]

Using the thermodynamic relation between $q$ and entropy and pressure perturbations the kernel ${\cal K}_{\rm s}$ can be expressed as
\begin{align*}
{\cal K}_{\rm s} & = -p\,\overline{\sigma\,q} \\
 & = \texttt{A}_4\,p\,\overline{\sigma\,p} + \texttt{A}_7\,p\,\overline{\sigma\,s}.
\end{align*}

Thus the work integral separates into two contributions
\begin{equation}
	\WS = \WC + \WT,
	\label{eq:Ws}
\end{equation}
where 
\[
\WC = \int_{x_b}^{x_{\rm s}} C(x) \texttt{A}_4\Re\left(p\,\overline{\sigma p}\right)\,dx
\]
and
\[
\WT = \int_{x_b}^{x_{\rm s}} C(x) \texttt{A}_7\Re\left(p\,\overline{\sigma s}\right)\,dx.
\]
The term $\WC$ represents the pressure/compressional part, and the term $\WT$ the pressure-entropy/thermal part.

\subsection{Derivation of the norm integral and surface terms}
For radial pulsation the linearized continuity equation may be rewritten in the following form, exposing $q$
\[
q=d'+3\,d,
\]
which, when substituted into \mEq{Kay} leads to
\[
{\cal K}_{\rm s} = -p\,\overline{\sigma (d'+3\,d)}.
\]

Substituting into $\WS$ it is obtained that
\[
\WS = -\Re{\int C\,p\,\overline{\sigma d'}\, dx}-3\Re{\int C\,p\,\overline{\sigma d}\, dx}.
\]

Integrating the first term by parts gives
\[
-\int C\,p\,\overline{\sigma d'}\,dx=-\left[C\,p\,\overline{\sigma d}\right]_{x_b}^{x_{\rm s}}+\int \left(Cp\right)'\overline{\sigma d}\,dx.
\]

Hence 
\[
\WS = -\Re{\left[C\,p\,\overline{\sigma d}\right]_{x_b}^{x_{\rm s}}} + \Re{\int \left[\left(Cp\right)'-3\,C\,p\right]\overline{\sigma d}\,dx}.
\]

Now
\[
\left(Cp\right)' - 3Cp = C'p+Cp'-3Cp.
\]

Since
\[
C\propto r^3\,P,
\]
we have
\[
\frac{C'}{C}=3+\frac{d \ln{P}}{dx}=3-\texttt{A}_3,
\]
using the definition of $\texttt{A}_3$.

Thus
\[
C'=(3-\texttt{A}_3)\,C.
\]
So
\[
\left(C\,p\right)'-3\,C\,p=(3-\texttt{A}_3)\,C\,p+C\,p'-3\,C\,p=C\left(p'-\texttt{A}_3 p\right).
\]

Therefore
\[
\WS = -\Re{\left[Cp\overline{\sigma d}\right]_{x_b}^{x_{\rm s}}}+\Re{\int C\left(p'-\texttt{A}_3p\right)\overline{\sigma d}\,dx}.
\]

From the radial pulsation equation for pressure perturbation it follows that
\[
p'-\texttt{A}_3p=\texttt{A}_3\left(4-\texttt{A}_2\sigma^2\right)d.
\]

Hence
\[
\WS = -\Re{\left[Cp\,\overline{\sigma\,d}\right]_{x_b}^{x_{\rm s}}}+\Re{\int C\texttt{A}_3\left(4-\texttt{A}_2\sigma^2\right)d\,\overline{\sigma\,d}\,dx}.
\]

Since $C$, and \texttt{A}'s are real, then
\[
\Re{\left[(4-\texttt{A}_2\sigma^2)d\overline{\sigma d}\right]}=\Re{(\sigma)}\left(4-\texttt{A}_2|\sigma|^2\right)|d|^2.
\]

This suggests to define the boundary term
\begin{equation}
\Phi(x)=C(x)\Re{\left( p(x)\,\overline{\sigma\,d(x)}\right)}
\label{eq:Phi}
\end{equation}
and the closure functional
\[
\NS=\int_{x_b}^{x_{\rm s}} C(x) \texttt{A}_3\left(4-\texttt{A}_2|\sigma|^2\right)|d|^2\,dx.
\]	

We may further expand the formula for $\NS$ into two terms as
\begin{equation}
\NS = \NC + \NT,
\label{eq:Ns}
\end{equation}
where
\[
\NC = -\int_{x_b}^{x_{\rm s}} C(x)\texttt{A}_3\texttt{A}_2|\sigma|^2|d|^2\,dx,
\]
and
\[
\NT = 4\int_{x_b}^{x_{\rm s}} C(x) \texttt{A}_3|d|^2\,dx.
\]
The term $\NC$ is the inertial contribution to the norm $\NS$, and is most directly associated with the inertial kinetic energy due to dependence on $|\sigma|^2|d|^2$. However the form of $\NS$ as a whole is induced by the quadratic balance relation and $\NC$ is not the classical kinetic energy integral. The second term $\NT$ is frequency independent and arises from the background stratification and spherical geometry in the radial pressure equation. It may be considered as an acoustic structure contribution to the norm $\NS$.

It should be noted that contrary to classical kinetic energy ($EK$) integral the functional $\NS$ is not positive definite.

\subsection{The quadratic balance relation}

The above derivation leads to the relation
\begin{equation}
	\WS+\PHS-\PHB=\Re{(\sigma)}\,\NS.
\label{eq:BalanceEq}
\end{equation}
which is formally analogous to the classical relation between work and energy integrals. Here $\WS$ is the work integral, and $\NS$ may be viewed as a generalization of kinetic energy integral, though it is a closure functional required by this equation, while the $\Phi$ terms are the surface terms.

Using \mEq{BalanceEq} it is possible to introduce the often used check for the consistency of the determination of the excitation rate by comparing 
\begin{equation}
\RES_{\rm check} = \frac{\WS+\PHS-\PHB}{\NS}
\label{eq:checkRe}
\end{equation}

to the real part of the eigenfrequency obtained from the solution of the boundary value problem. In this role \mEq{checkRe} plays the same role as classical consistency check, although the formulae used to compute the check differ substantially from their classical counterparts.

In addition to enabling the computation of $\RES_{\rm check}$ the $\WS$ and $\NS$ terms may be used to obtain information about the driving and damping of oscillations, as well as of the relative contributions to $\WS$ and $\NS$ from the two constituent terms that each of these quantities is composed of. Thus $\WS$ may be used to determine driving and damping regions and to examine the contribution to driving from the compressional and thermal parts.

The surface terms (\mEq{Phi}) are of importance for they may be used to compare the relative contribution to the balance of the surface and volume terms. If any of these surface terms becomes large as compared to $\WS$, the volume integral, it means the conditions at the corresponding surface cause concentration of pulsation wave flux there. An effect that should not become dominant under expected behavior of outer and inner boundary conditions. We use these two surface terms regularly to check for signs of problems with boundary conditions.

\subsection{Comparison to EK and ED integrals}

It is useful to compare the quantities introduced in the preceding section with classical kinetic energy and dissipation integrals. The quadratic forms for these integrals in linear pulsation theory are discussed in Unno et al. (1989). Equating the average over cycle of the dissipated energy with the change of pulsation energy leads to the well established relation between dissipation integral, the kinetic energy integral and the mode excitation rate.

The present formulation differs from the conventional cycle average ED-EK construction, although it represents the same underlying mechanical balance as the enseble-average formulation of Glatzel (1994). Instead of deriving second order forms of average work done or kinetic energy and postulating the equality of their averaged integrals we start from the sesquilinear kernel (${\cal K}_{\rm s}$) and use the linear nonadiabatic radial pulsation equations to obtain the relations. This gives a consistent derivation of the forms for $\WS$, $\NS$ and surface terms. The surface terms in \mEq{BalanceEq}, which are often omitted (though see eg. Osaki, 1977) are included in our balance equation as they are part of the consistent formulation. The quantities obtained by us, while they form a relation similar to the classical equation, are not derived from the same assumptions as $ED$ and $EK$. For this reason it is useful to compare ($ED$,$EK$) with ($\WS$,$\NS$). 

The formulae for the kinetic energy and dissipation integral used for the comparison will be those given in Dziembowski (1977, 1994). Physical discussion of the terms appearing in the dissipation integral may be found in Dziembowski (1994) and Pamyatnykh (1999), and a broader discussion of the dissipation integral and kinetic energy integrals may be found in Cox (1980) and Unno et al. (1989).

For weakly nonadiabatic pulsations it may be expected that both approaches will lead to similar results, while for highly nonadiabatic pulsations in supergiant envelopes the differences may become substantial, and they may reveal parts that are hidden or lost in the classical interpretation.

The differences and similarities may be directly seen by comparing the derivations of the components of $\WS=\WC+\WT$
with the formula for the work integral $W$ in Pamyatnykh (1999). It may be
shown that for the radial pulsation case considered here the
\[
W \sim -\Re{(\overline{p}\,\omega\,s)} \sim -\Re{(\overline{p}\,\sigma\,s)}
\]
and it corresponds to the term $\WT$.

The term $\WC$ is absent in the formula for $W$ because the formula for $W$ is derived
for near-adiabatic cases where it is assumed that $\sigma$ is imaginary, or at least $|\RES|\ll|\IMS|$,
and since $\WC$ depends on $|p|^2\,\Re{(\sigma)}$ then for nearly-adiabatic modes with $\sigma\approx i\,\sigma_{\rm I}$ a term like $\WC$ would be neglected in the derivation of $W$. But for strongly nonadiabatic modes, with $|\RES|\sim|\IMS|$ the term $\WC$ cannot be neglected.

By analyzing the form of $\NS=\NC+\NT$ it is seen that $\NC$ is the closest analogue of the
kinetic energy integral though, as it may be seen in Dziembowski (1977) from the definition of $EK=\omega_{\rm I}^2\int |\overrightarrow{\xi}|^2\rho\, d^3\vec{x}$ given there (see also Osaki, 1977), that
in $EK$ there appears the imaginary part of the frequency squared, while $\NC$ depends on the $|\sigma|^2$. The presented derivation clarifies why the natural nonadiabatic generalization $\NC$ involves $|\sigma|^2$ rather than only $\sigma_{\rm I}^2$.

The term $\NT$ is absent in the definition of $EK$ and it causes the $\NS$ to become not positive definite, and thus may have substantial impact on the check condition \mEq{checkRe}. This term is not dependent on $\sigma$, thus it exists even in case of purely adiabatic modes. Hence, contrary to $W_s$, the $N_s$ is not reducible to the adiabatic kinetic energy integral for purely adiabatic pulsation.

The classical $EK$ and $ED$ relate to our $\NS$ and $\WS$ as follows
\begin{align}
\begin{alignedat}{3}
 \WS &{}= \WC + \WT &\;&{}= \WC - ED, \\
 \NS &{}= \NC+\NT &\;&{}= -2\left(\frac{|\sigma|}{\IMS}\right)^2\, EK +\NT.
\end{alignedat}
\label{eq:EKED}
\end{align}

It follows from these equations that in order to compare the effects of the derivation presented here it is sufficient to analyze the impact of the $\WC$ and $\NT$ terms on $\WS$ and $\NS$. 

From the above it follows that the commonly used estimate based on the ratio of $ED/(2\,EK)$ should therefore not be expected to reproduce real part of a nonadiabatic eigenfrequency, at least not for sufficiently nonadiabatic modes.

The behavior of ($\WS$,$\NS$) and ($ED$,$EK$) will be compared and analyzed in the following section. We will assume the same form of the surface terms in both cases to enable comparison.

\section{Comparison of classical and quadratic-balance diagnostics}

In this section we compare the behavior of the terms entering into the work and the generalized norm integrals and compare the results to those obtained using the classical $EK$ and $ED$ for a selection of radial pulsation modes in supergiant envelopes. 

\subsection{Work integral $W_s$}

\begin{figure}[htb]
	\includegraphics{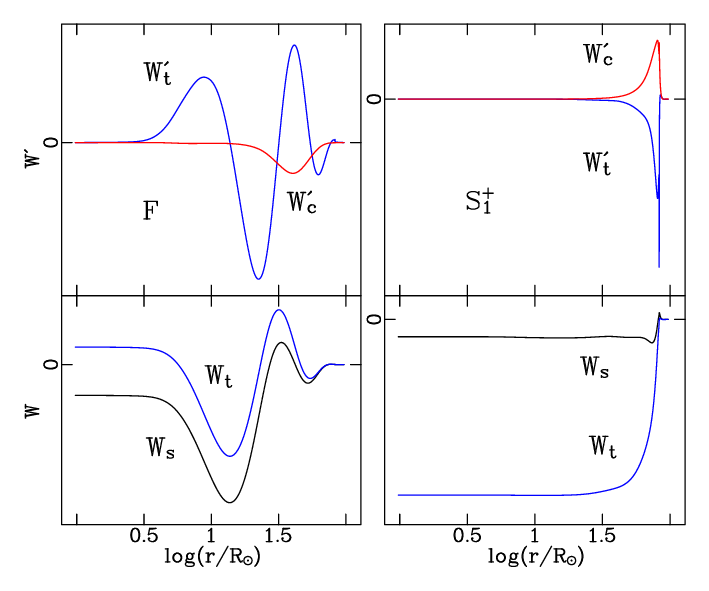}
	\FigCap{The integrands of $\WC$ and $\WT$, denoted as $\WC'$ and $\WT'$ are plotted in the upper panel, and the cumulative integrals for $\WS$ and $\WT$ in the lower panel for a fundamental $F$ mode on the left and strange $S^{+}_1$ mode on the right. The integrals are computed from the surface downwards. It is seen that the compressional term $\WC$ plays a significant role, as $\WS\neq\WT$.}
	\label{fig:Fig1}
\end{figure} 

The work integral $W_s$, \mEq{Ws}, when derived from the kernel \mEq{Kay} is found to be composed of two terms with the thermal part $W_t$ corresponding to the negative of the classic dissipation integral $ED$, and an additional compressional term $W_c$. In what follows we will analyze how this term affects the work integral $W_s$ for acoustic modes in the envelope of a supergiant star for the case of ordinary p-modes and for strange modes.   

We selected two pulsation modes - one corresponding to the fundamental mode and one to the lowest frequency excited strange mode $S^+_1$ for the model with $\log(T_{\rm eff})=3.8$, $\log(L/L_{\odot})=4$ and $M/M_{\odot}=0.69$. To help with the analysis we employ the boundary conditions $(3,4)\text{--}(1,2)$ to avoid large surface terms for strange modes. The outer selector $(3,4)$ helps avoid large surface values of $\PHS$, while the inner selector $(1,2)$ prevents large values of $\PHB$.  For this form of boundary conditions Zalewski (2026a) found a clear radial mode spectrum in which strange modes form nearly symmetric pairs of $\pm\RES$ and similar $\IMS$. 

In \mFig{Fig1} both the integrands and the work integrals are shown for these two modes. As may be seen from the upper panels in \mFig{Fig1} the integrand of the compressional term - $\WC'$ has smaller values relative to the $\WT'$ for the fundamental mode, however $\WC$ has substantial impact on $\WS$ as the sign of the integrated $\WS$ is opposite to what would be obtained from the term $\WT$ alone. Similarly for the strange mode the term $\WC$ results in the work integral being substantially different from what would be obtained using only the dissipation-integral-like term $\WT$. Thus, it turns out, the term $\WC$ has substantial impact on $\WS$ in both cases. 
\begin{figure}[htb]
	\includegraphics{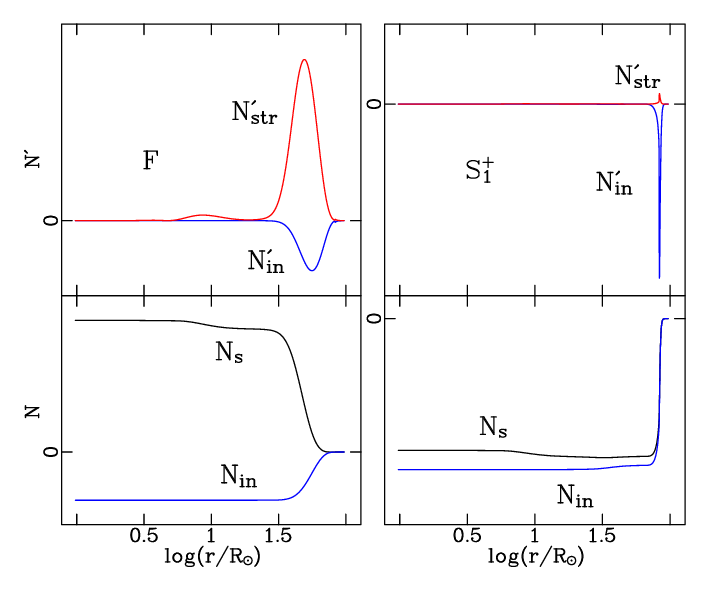}
	\FigCap{The integrands of $\NC$ and $\NT$, denoted by $\NC'$ and $\NT'$ respectively, are plotted in the upper panel, and the cumulative integrals for $\NS$ and $\NC$ in the lower panel for a fundamental $F$ mode on the left and strange $S^{+}_1$ mode on the right. It is seen that the term $\NT$ plays a role for both types of modes, as $\NS\ne\NC$.}
	\label{fig:Fig2}
\end{figure} 

\subsection{Generalized norm integral \(\NS\)}

Similarly to the work integral, the generalized norm integral $\NS$ may be split into the inertial contribution $\NC$ and the structural contribution $\NT$.

The plots of $\NC'$ and $\NT'$, the integrands of $\NC$ and $\NT$, as well as of the cumulative integrals $\NC$ and $\NT$ are shown in \mFig{Fig2}. 

For the fundamental mode, $\NT$ provides the dominant contribution to $\NS$. For the strange mode $S^{+}_1$, $\NT$ is non-negligible but remains smaller than the inertial contribution $\NC$. Thus the relative importance of $\NT$ differs substantially between the two types of acoustic modes. It may be expected that inclusion of $\NT$ and $\WC$ terms in the balance equation may affect the balance and the value of the $\RES_{\rm check}$.

\subsection{Magnitude of the boundary terms}

In addition to the work integral and the norm integral, \mEq{BalanceEq} for the balance check contains the surface terms $\Phi$. Surface terms appear also in the classical derivation of kinetic energy and dissipation integral. Unno et al. (1989) provide a discussion of their origin and meaning.

These terms are often neglected using arguments of vanishing pressure at the surface or continuity at the center. See however Osaki (1977) who neglects the outer surface term by assuming zero pressure above the surface, but retains the inner boundary's surface term to account for energy transmitted towards the interior. On the other hand Shibahashi (2005), while also neglecting the outer surface term, discusses conditions which could lead to its inclusion and the impact it could have on pulsation modes trapped below the outer surface. 

Most often the surface terms are ignored and the discussion is focused on either the estimation of excitation rate from $EK$ and $ED$ or on analysis of the work integral to determine driving regions (see eg. Dziembowski, 1994, Pamyatnykh, 1999).

In case of acoustic modes in the envelopes of supergiant stars there is no well defined surface above which pressure would vanish, nor do the radial modes reach the stellar interior. In fact both boundaries are regions of gradual transitions to the exterior atmosphere or the deeper envelope regions. Thus it is not possible to assume zero pressure or similar approximations and the surface terms need be included. 

For the form of the boundary conditions selector $(3,4)$\text{--}$(1,2)$ chosen here the surface terms are small, as will be seen from \mTab{Tab1} below, so their role in the balance equation is not obvious, however for some of the other types of selectors, particularly those that admit large outgoing acoustic flux the surface terms can become comparable in magnitude to $\WS$. The surface terms are not an arbitrary assumption in our formulation, but need to be present for the closure of the balance equation \mEq{BalanceEq} and provide important information about boundary conditions. 

\subsection{Comparison of balance equation terms}

\MakeTable{lllllll}{12.5cm}{Contribution to balance check from the volume integral and surface terms plus the relative difference $\epsilon_{\rm bal}=1-\RES_{\rm check}/\RES$ between the computed $\RES_{\rm check}$ and the real part of the eigenfrequency for selected p-modes, strange modes (S) and a thermal mode (T). For boundary conditions selector $(3,4)\text{--}(1,2)$. \label{tab:Tab1}}
{
	\hline
	  Mode
	& $\IMS$
	& $\phantom{-}\RES$
	& $\phantom{-}\epsilon_{\rm bal}$
	& $\phantom{-}\WS/\NS$
	& $\phantom{-}\PHS/\NS$
	& $\phantom{-}\PHB/\NS$ \\
	
	\hline
	
$F^{\phantom{X^X}}$       & $ \phantom{1}1.61$  & $           -5.38{\times}10^{-3}$ & $ \phantom{-}5.4{\times}10^{-6}$  & $           -5.29{\times}10^{-3}$  & $           -7.90{\times}10^{-6}$  & $ \phantom{-}7.77{\times}10^{-5}$ \\
$1$-ov                    & $ \phantom{1}2.79$  & $           -4.70{\times}10^{-2}$ & $           -1.7{\times}10^{-6}$  & $           -4.58{\times}10^{-2}$  & $           -3.03{\times}10^{-6}$  & $ \phantom{-}1.18{\times}10^{-3}$ \\
$16$-ov                   & $           19.32$  & $           -2.42{\times}10^{-1}$ & $           -4.1{\times}10^{-6}$  & $           -2.36{\times}10^{-1}$  & $           -1.04{\times}10^{-5}$  & $ \phantom{-}5.72{\times}10^{-3}$ \\

	\hline
	
$S^{+\phantom{X^X}}_1$    & $ \phantom{1}4.66$  & $ \phantom{-}3.60               $ & $           -5.6{\times}10^{-7}$  & $ \phantom{-}3.60               $  & $           -3.10{\times}10^{-3}$  & $ \phantom{-}2.62{\times}10^{-5}$ \\
$S^{-}_1$                 & $ \phantom{1}4.76$  & $           -3.62               $ & $           -2.8{\times}10^{-7}$  & $           -3.63               $  & $ \phantom{-}3.20{\times}10^{-3}$  & $           -2.10{\times}10^{-4}$ \\
$S^{+}_2$                 & $ \phantom{1}8.87$  & $ \phantom{-}3.47               $ & $           -2.3{\times}10^{-6}$  & $ \phantom{-}3.47               $  & $           -5.24{\times}10^{-3}$  & $           -2.71{\times}10^{-6}$ \\
$S^{+}_3$                 & $           13.19$  & $ \phantom{-}3.51               $ & $           -2.3{\times}10^{-6}$  & $ \phantom{-}3.52               $  & $           -3.61{\times}10^{-3}$  & $           -4.31{\times}10^{-6}$ \\
$S^{+}_4$                 & $           17.49$  & $ \phantom{-}3.58               $ & $           -1.3{\times}10^{-5}$  & $ \phantom{-}3.58               $  & $           -1.44{\times}10^{-3}$  & $           -2.22{\times}10^{-5}$ \\
	
	\hline
	
$T^{+\phantom{X^X}}_1$    & $ \phantom{1}0.00$  & $ \phantom{-}2.27               $ & $           -4.4{\times}10^{-7}$  & $ \phantom{-}2.27               $  & $           -5.19{\times}10^{-9}$  & $           -2.23{\times}10^{-3}$ \\

	\hline
	
}

By dividing the quantities entering the equation for $\RES_{\rm check}$ in \mEq{checkRe} by $\NS$ we obtain three numbers representing contributions to the real part of frequency $\sigma$ coming from the volume integral and the two surface terms. This makes it possible to inspect their relative roles as well as examine how they behave for different modes. The results are given in \mTab{Tab1}, the data are for the same model as in \mFig{Fig1}. The modes were computed using the continuous renormalization integrator $QC$ with tracking transformation and with variable step size denoted as "phase-controlled" in Zalewski (2026b). However the integrals $\WS$ and $\NS$ were computed using simple quadratures based on tabulated values of eigenfunctions.

From \mTab{Tab1} it follows that for all modes listed the relative difference in the real part of eigenfrequency $|\epsilon_{\rm bal}|\lesssim 10^{-5}$ with substantially better accuracy for low frequency modes, which may suggest that the integrals should be computed using a more accurate integration method.

For all types of modes listed in \mTab{Tab1} the contribution of the surface terms is small. Thus for the assumed form of the boundary conditions selector of $(3,4)\text{--}(1,2)$ the surface terms do not have a substantial effect on the balance.

\subsection{Effect of omitting the additional balance terms}
We assess the importance of the terms $\WC$ and $\NT$ and their effect on the excitation rate by comparing with the results that would be obtained from classical formulation.

We thus introduce, see \mEq{EKED}, a 
\[
W^*_{\rm s}= \WT= -ED,
\]
and
\[
N^*_{\rm s}= \NC= -2\left(\frac{|\sigma|}{\IMS}\right)^2EK.
\]

\MakeTable{llll}{12.5cm}{Comparison of the excitation rate check values obtained using the truncated balance relation, \mEq{TruncEq}, for the same modes as in \mTab{Tab1}.\label{tab:Tab2}}
{
	\hline
	Mode
	& $\IMS$
	& $\phantom{-}\RES$
	& $\phantom{-}\RES_{\rm trunc}$ \\
	
	\hline
	
	$F^{\phantom{X^X}}$       & $ \phantom{1}1.61$  & $           -5.38{\times}10^{-3}$  & $           -8.3{\times}10^{-3}$   \\
	$1$-ov                    & $ \phantom{1}2.79$  & $           -4.70{\times}10^{-2}$  & $ \phantom{-}1.0{\times}10^{-1}$   \\
	$16$-ov                   & $           19.32$  & $           -2.42{\times}10^{-1}$  & $ \phantom{-}1.1{\times}10^{-1}$   \\
	
	\hline
	
	$S^{+\phantom{X^X}}_1$    & $ \phantom{1}4.66$  & $ \phantom{-}3.60               $  & $ \phantom{-}31.8              $  \\
	$S^{-}_1$                 & $ \phantom{1}4.76$  & $           -3.62               $  & $           -32.9              $  \\
	$S^{+}_2$                 & $ \phantom{1}8.87$  & $ \phantom{-}3.47               $  & $ \phantom{-}42.3              $  \\
	$S^{+}_3$                 & $           13.19$  & $ \phantom{-}3.51               $  & $ \phantom{-}52.4              $  \\
	$S^{+}_4$                 & $           17.49$  & $ \phantom{-}3.58               $  & $ \phantom{-}61.8              $  \\
	
	\hline
	
	$T^{+\phantom{X^X}}_1$    & $ \phantom{1}0.00$  & $ \phantom{-}2.27               $  & $           -136.0             $  \\
	
	\hline
	
}

We use $\NC$ rather than $EK$ because using classical kinetic energy integral would introduce an obvious bias to the result. We also retain the computed surface terms $\Phi$ in the calculation of the excitation rate. Hence, in what follows we check the effects of omitting the new terms $\NT$ and $\WC$. Thus, from \mEq{checkRe}, upon substitution it is obtained that
\begin{equation}
\RES_{\rm trunc}=\frac{W^*_{\rm s}+\PHS-\PHB}{N^*_{\rm s}}.
\label{eq:TruncEq}
\end{equation}
The results for $\RES_{\rm trunc}$ are given in \mTab{Tab2} for the same modes as listed in \mTab{Tab1}. 

By comparing the $\RES_{\rm trunc}$ to $\RES$ for selected modes of the model it is seen that the omission of the terms $\NT$ and $\WC$ in Eqs.~(\ref{eq:Ws}) and (\ref{eq:Ns}) leads to substantial disagreement of the check with the real part of eigenfrequency.

It follows that all the terms entering the balance equation \mEq{checkRe} need to be retained in order to obtain a proper check condition on the excitation rate. In order to obtain a balance the terms $\NT$ and $\WC$ cannot be neglected.

\section{Diagnostics from balance relation for ordinary and strange modes}

In this section we will examine the diagnostics provided by the balance relation for radial ordinary and strange modes in the AGB envelopes. We use the same model and boundary selector as in the previous section. 

\begin{figure}[htb]
	\includegraphics{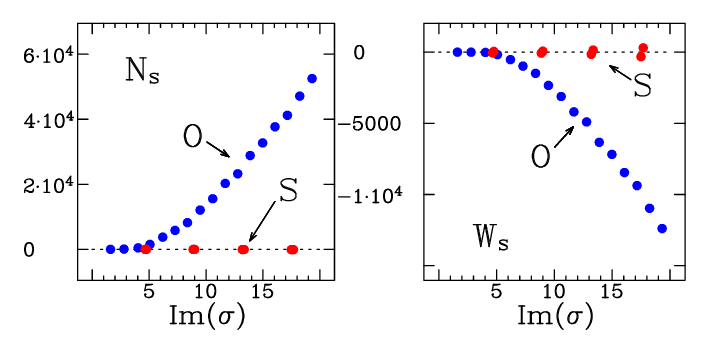}
	\FigCap{Generalized norm integral $\NS$ (left panel) and the work integral $\WS$ (right panel) as functions of pulsation frequency $\IMS$. Values for ordinary modes are marked in blue, while those for strange modes in red. }
	\label{fig:Fig3}
\end{figure} 

In \mFig{Fig3} the $\NS$ and $\WS$ are plotted for ordinary and strange modes as a function of $\IMS$. It may be seen that $\NS$ for ordinary modes increases with increasing frequency and its value for ordinary modes is much larger than for strange modes. It should also be noted that for ordinary modes the $\NS$ is positive while the contribution from the kinetic energy analogue $\NC$  is negative, thus the norm integral is determined by the term $\NT$ introduced in \mEq{Ns}. The graph for $\WS$ in the right panel shows a similar distinction between ordinary and strange modes. For ordinary modes $\WS$ is negative and decreases with the increase of frequency, whereas its values for strange modes are much smaller.

Thus, both the generalized norm integral and the work integrals show markedly different behavior for ordinary and strange modes. It is therefore useful to examine their properties in detail.

\subsection{Frequency dependence of the generalized norm integral}

\begin{figure}[htb]
	\includegraphics{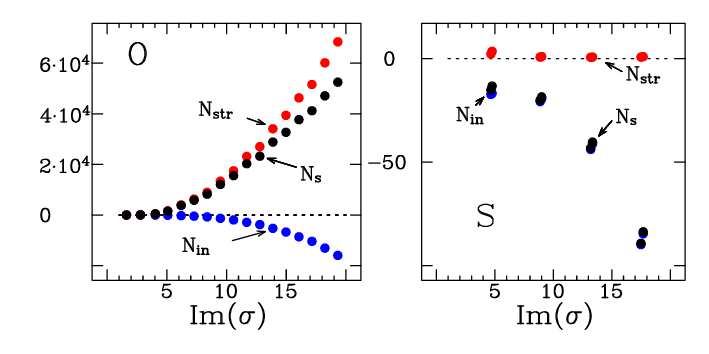}
	\FigCap{The dependence of $\NS$ (black points), $\NC$ (blue points) and $\NT$ (red points) on $\IMS$ is shown for the ordinary (O) modes (left panel) and for the strange (S) modes (right panel). For ordinary modes $\NS\sim \NT$ while for strange modes $\NS\sim \NC$. }
	\label{fig:Fig4}
\end{figure} 

In \mFig{Fig4} the dependence of $\NS$ and its components on $\IMS$ is shown for ordinary modes in the left, and strange modes in the right panels. The component $\NC$, which depends on the inertial contribution of the mode, proportional to ($|\sigma\,d|^2$) is negative and decreasing with increase of pulsation frequency for both the ordinary and the strange mode. The component $\NT$ is positive and increases with frequency for ordinary modes, while it has much smaller values for strange modes. The values of $\NS$ and its components are three orders of magnitude smaller for the strange modes than for ordinary modes, as was already visible in the left panel in \mFig{Fig3}. 

For ordinary modes $\NS$ is determined mainly by the structural term $\NT$. For strange modes the situation is different: the structural term is small and $\NS$ follows the inertial term $\NC$. Thus the same balance relation is satisfied in the two cases in a different way. In the ordinary modes the dominant part of $\NS$ is a term which has no analogue in the classical kinetic-energy integral, whereas in the strange modes the dominant part is the kinetic-energy-like term. This is why $\NS$ should not be interpreted as a simple replacement of the classical kinetic energy.

\subsection{Frequency dependence of the work integral}
\begin{figure}[htb]
	\includegraphics{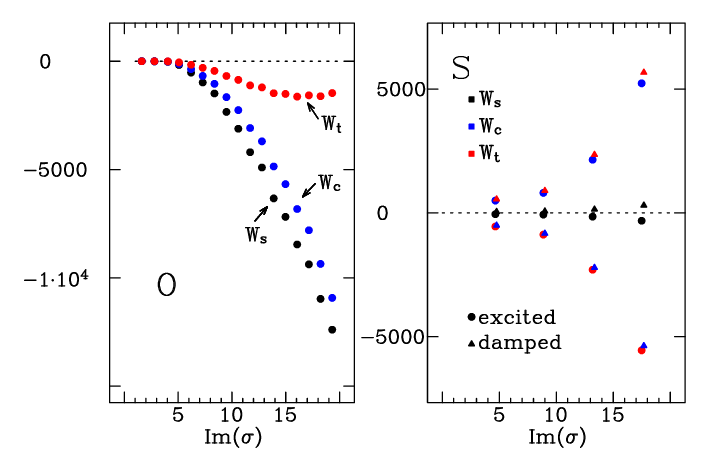}
	\FigCap{The dependence of $\WS$ (black points), $\WC$ (blue points) and $\WT$ (red points) on $\IMS$ is shown for the ordinary (O) modes (left panel) and for the strange (S) modes (right panel). For ordinary modes $\WS\sim \WC$ and its magnitude increases with frequency, while for strange modes $\WS$ is close to zero with $\WC$ having opposite sign to $\WT$ for a given strange mode. }
	\label{fig:Fig5}
\end{figure} 

As in the case of $\NS$ the behavior of $\WS$ for ordinary and strange modes differs, as may already be seen in \mFig{Fig3}. It is therefore useful to examine the $\WS$ and its components as a function of pulsation frequency. The plots for $\WS$, $\WC$ and $\WT$ are shown in \mFig{Fig5}, for ordinary modes in the left panel and for strange modes in the right one.

It is seen from \mFig{Fig5} that, for ordinary modes, the work integral $\WS$ follows mainly the compressional part $\WC$, while the part $\WT$, associated with classical dissipation integral is much smaller. Thus, similarly to the case with generalized norm integral, the dominant contribution for ordinary mode comes from a term that is not included in classical work integral expression. This indicates that the explicitly nonadiabatic contribution represented by $\WC$, which is proportional to ($\RES\,|p|^2$), plays an important role in the work integral of ordinary modes in AGB envelopes. It also implies that, in supergiant envelopes, a diagnostic based only on the classical $p\overline{\sigma s}$ term may miss contributions that are essential for the modal work balance.

For strange modes the situation is different. The terms $\WC$ and $\WT$ are not small, their magnitudes are comparable to the magnitude of the dominant term for ordinary modes. However for strange modes these terms are of opposite sign, and thus their sum $\WS$ is much smaller. This leads to the $|\WS|$ for strange modes being an order of magnitude smaller than for ordinary ones. The $\WS$ is positive for one of the strange modes in a pair and negative for the other.

The strange mode pairs also show an approximate interchange of the two contributions. The compressional contribution for one mode in the pair is close to the thermal contribution for the other member, while for each individual strange mode the two terms nearly compensate each other. This suggests that the work integrals for two members in a strange mode pair result from a balance of two opposing contributions - viz. the compressional term $\WC$ and the thermal term $\WT$. In this sense the compressional term acts in opposition to the traditional pressure-entropy work term ($p\overline{\sigma\,s}$) in the work balance of a strange mode. 

Thus, in contrast to the generalized norm integral $\NS$, which for strange modes was determined primarily by the inertia term $\NC$, the work integral $\WS$ for these modes is determined by a near cancellation of two opposing contributions from the terms $\WC$ and $\WT$. 

It should be emphasized that the sign of $\WS$ does not in itself determine whether the mode is excited or damped. The excitation rate, as given in \mEq{checkRe}, is determined in conjunction with the sign of $\NS$ (which for strange modes has negative values), and by the surface terms (which are close to zero for the assumed boundary conditions).

\subsection{Interpretation of balance diagnostics for ordinary modes}

In order to understand why the terms $\NT$ and $\WC$ determine the values of generalized norm integral $\NS$ and work integral $\WS$ for ordinary modes of radial pulsation in AGB envelopes it is necessary to analyze the component integrands.

For the work integral the two components may be written schematically as
\begin{align*}
\begin{alignedat}{3}
\WC' &{}= C(x)\,\texttt{A}_4 \, \Re(\sigma)\,|p|^2 &\;&{}= C_{1}(x) \, \Re(\sigma)\,|p|^2, \\
\WT' &{}= C(x)\,\texttt{A}_7 \, \Re(p\overline{\sigma\,s}) &\;&{}= -C_2(x) \, \Re(p\overline{\sigma\,s}),
\end{alignedat}
\end{align*}
where the weight factors satisfy $C_1>0$ and $C_2>0$ in the envelope. The quadratic amplitude factor in $\WC'$ has a fixed sign, determined by the sign of $\Re(\sigma)$. Thus $\WC$ may be written as 
\[
\WC=\Re(\sigma)\,{\WC}_{0},
\]
where the ${\WC}_{0}>0$. 

The term $\WT'$, has a negative sign and positive weight factor, but the sign of the quadratic pressure-entropy-rate term may vary between the driving and damping regions in the envelope. Thus $\WT$ may be positive or negative. For the ordinary modes considered here, the quadratic pressure term entering $\WC$ is much larger than the amplitude of the quadratic term in $\WT$, mainly because $|p|\gg|s|$. Therefore $|\WC|>|\WT|$. This explains why, for ordinary radial modes in the AGB envelopes considered here, the sign of $\WS$ is essentially  determined by the $\Re(\sigma)$, as seen in \mFig{Fig5}. 

The generalized norm integral $\NS$ is a sum of two terms, one of which ($\NC$) is negative. The weight factors in both the $\NC$ and $\NT$ integrals are positive and both integrands contain the same quadratic displacement perturbation amplitude, $|d|^2$. Thus which of these terms dominates $\NS$ is determined by the shape of the displacement perturbation relative to the two weights. Denoting by $D_{\rm in}(x)$ and $D_{\rm str}(x)$ the weight factors for $\NC$ and $\NT$ respectively, we have schematically
\begin{align*}
	\begin{alignedat}{3}
\NC' &{}= -&\;&{}D_{\rm in}(x) |d|^2, \\
\NT' &{}= &\;&{}D_{\rm str}(x) |d|^2.
\end{alignedat}
\end{align*}

\begin{figure}[htb]
	\includegraphics{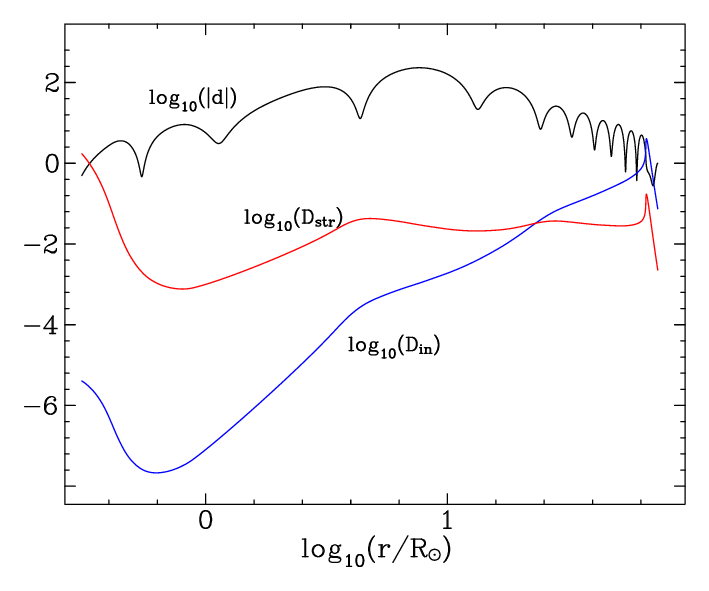}
	\FigCap{Comparison of the $\log(D_{\rm in})$, $\log(D_{\rm str})$ and $\log(|d|)$ for the sixth radial overtone in the envelope of the model with $\log(T_{\rm eff})=3.8$. In the regions (below ionization zones), where $|d|$ is large, $D_{\rm str}$ is larger than $D_{\rm in}$. }
	\label{fig:Fig6}
\end{figure} 

From \mFig{Fig6} it is seen that in regions where $D_{\rm str}>D_{\rm in}$, the displacement amplitude is large, and the contribution from the term $\NT$ dominates $\NS$. For the fundamental mode, the maximum of $D_{\rm in}$ in the H/He~I ionization zone becomes approximately comparable to that of $D_{\rm str}$. For higher overtone modes the $D_{\rm in}$ dominates in the ionization region because of the factor $|\sigma|^2$. At the same time, however, the amplitude of radial overtones decreases rapidly above the H/He~I ionization region, as may be seen from \mFig{Fig6}. As a result, for ordinary radial modes considered here the structural contribution $D_{\rm str}$ dominates and $\NS\approx \NT>0$.

\subsection{Interpretation of balance diagnostics for strange modes}

In this section we analyze the factors affecting the $\WS$ and $\NS$ for strange modes. We focus specifically on strange modes obtained using the \((3,4)\) boundary selection because for this form of the outer boundary selector a symmetrical spectrum of strange modes is obtained and the strange modes have largest excitation rates compared with other forms of boundary conditions analyzed in Zalewski (2026a). 

\begin{figure}[htb]
	\includegraphics{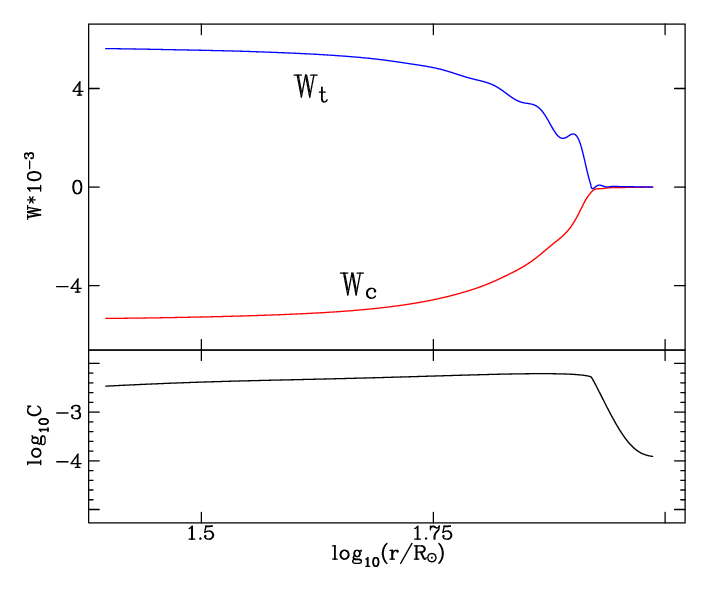}
	\FigCap{The contributions to the work integral - $\WC$ and $\WT$ shown in the regions of the envelope above He~II ionization for the strange mode $S^{-}_4$. Both $\WC$ and $\WT$ are integrated from the surface inwards. The two terms have opposite signs, show driving and damping in the H/He~I region and their values stabilize before reaching the He~II region. The lower panel shows the plot of the scaling factor $\log(C)$. $C(x)$ has a maximum in H/He~I zone in the region of opacity peak and drops towards surface.}
	\label{fig:Fig7}
\end{figure} 

To analyze $\WS$ and $\NS$ we will decompose the integrands of the components of $\NS$ into the amplitude of radius perturbation and weights $D_{\rm in}$ and $D_{\rm str}$, similarly as for ordinary modes, while for strange modes it is more useful to decompose the integrands of work integral components into
\begin{align*}
	\begin{alignedat}{3}
		\WC' &{}= C(x)\,\texttt{A}_4 \, \Re(\sigma)\,|p|^2 &\;&{}= C(x) \, K_{\rm c}, \\
		\WT' &{}= C(x)\,\texttt{A}_7 \, \Re(p\overline{\sigma\,s}) &\;&{}= C(x) \, K_{\rm t},
	\end{alignedat}
\end{align*}
and write the integrand of the work integral as
\begin{align*}
	\WS' = C(x)\,K_{\rm s},
\end{align*}
where $K_{\rm s}=K_{\rm c}+K_{\rm t}$.

The reason for such a form of the decomposition becomes obvious from \mFig{Fig7}, where the work integral components $\WC$ and $\WT$ are shown for a damped strange mode $S^{-}_{4}$.  Near the surface these integrals stay close to zero until the H ionization zone is reached. As may be seen, from the plot of $\log(C(x))$ in the lower panel, the weighting factor $C(x)$ becomes small above the H ionization zone in the outer layers of the envelope. In these regions the thermal time scale becomes small compared with the dynamical time scale. Glatzel (1994) argued that those outer layers of the envelope do not have enough heat capacity to contribute substantially to work integral.

\begin{figure}[htb]
	\includegraphics{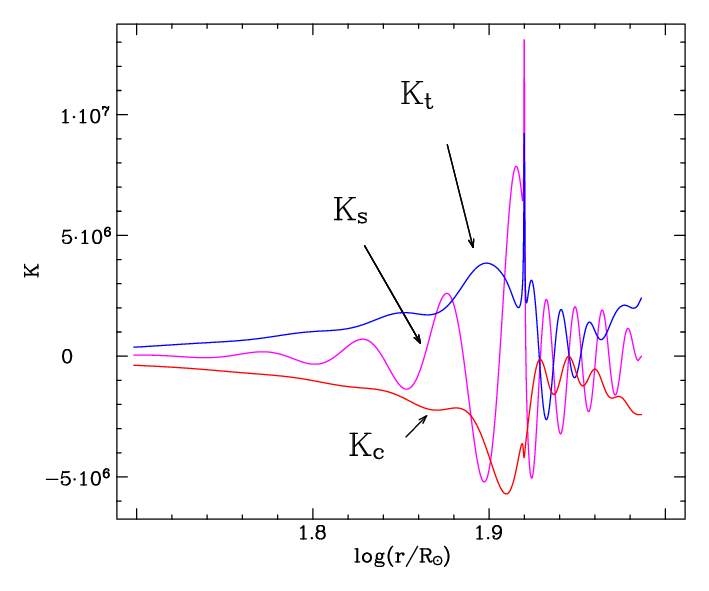}
	\FigCap{The reduced work kernels $K_{\rm c}$ and $K_{\rm t}$ become large and of opposite signs in the H and He~I zones for strange mode $S^{-}_4$. The sharp spike in $K_{\rm t}$ occurs in the H ionization region and coincides with the sharp spike in opacity. Above the H ionization zone both reduced work kernels show nonzero, oscillatory values but their effect on the work integral is cancelled by the rapid decrease of the scale factor $C(x)$ above H ionization zone. Also shown is the sum of the two kernels $K_{\rm s}=K_{\rm c}+K_{\rm t}$, which contrary to $K_{\rm c}$ and $K_{\rm t}$, alters sign in the H/He~I ionization zones.}
	\label{fig:Fig8}
\end{figure}

For the strange mode $S^{-}_4$ these outer regions are seen not to have much impact on $\WC$ or $\WT$ and it is only in the H and He~I ionization zones, where the $C(x)$ becomes large due to large opacity increase, that the integrals of these two quantities become large. For this strange mode they stabilize before the He~II ionization region is reached (which occurs at $x_{10}=\log_{10}(r/R_{\odot})\approx 1.4$ for this envelope model). Thus the driving and damping of this strange mode is confined to the H/He~I zones.
 
The effects of driving and damping may be seen clearly by analyzing the reduced work kernels $K_{\rm c}$ and $K_{\rm t}$. These are shown in \mFig{Fig8} for the same mode as in \mFig{Fig7} and for the same range of $x_{10}$. From \mFig{Fig8} it is seen that in the H/He~I ionization region the $K_{\rm c}<0$ while $K_{\rm t}>0$. The two kernels are of opposite signs and neither of them changes sign in the H/He~I ionization zone. Thus, for the mode $S^{-}_4$ the $\WC$ becomes negative in this region and $\WT$ becomes positive, as seen in \mFig{Fig7}. Because the sum of the two kernels, $K_{\rm s}$, is oscillatory and changes sign multiple times in the H/He~I ionization region it may be expected that the work integral $\WS$ based on $K_{\rm s}$ kernel will become small in magnitude compared with $\WC$ or $\WT$. It is thus necessary to examine the factors that lead to the nearly opposite behavior of $K_{\rm c}$ and $K_{\rm t}$ leading to large cancellation of their contributions.

The reduced kernels depend on real parts of the sesquilinear products of perturbation variables, as was introduced in the section on ordinary modes analysis, and
\begin{align*}
	K_{\rm c}&{}=\texttt{A}_4\,\Re(p\overline{\sigma\,p}),\\
	K_{\rm t}&{}=\texttt{A}_7\,\Re(p\overline{\sigma\,s}),
\end{align*}
where $\texttt{A}_4=\frac{1}{\Gamma_1}>0$ and $\texttt{A}_7=\left(\frac{\partial \ln \rho}{\partial \ln T}\right)_{\!P}<0$, thus the two factors preceding the products are of opposite sign. To determine the sign of the reduced kernel it is necessary to examine the signs of the sesquilinear products of corresponding perturbations.

\begin{figure}[htb]
	\includegraphics{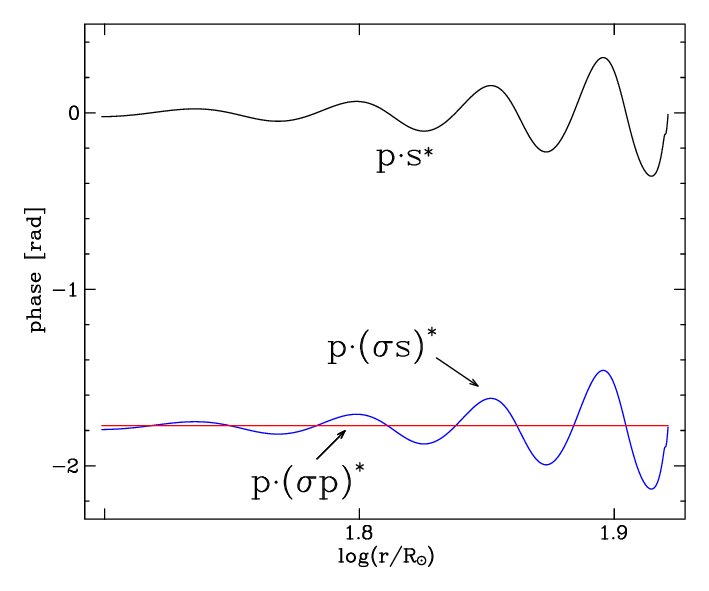}
	\FigCap{With the adopted outer boundary conditions of \((3,4)\) the phase difference between pressure ($p$) and entropy ($s$) perturbations in the region in and below H ionization zone is close to zero for the strange modes. The phase between $p$ and $s$ near the maximum of opacity has values $|\arg(p\overline{s})|<0.35$ and decreases inwards. Due to this effect the phases of the two sesquilinear products entering $K_{\rm c}$ and $K_{\rm t}$ become similar.}
	\label{fig:Fig9}
\end{figure} 

It is clear that $\Re(p\overline{\sigma\,p})=\Re(\sigma)|p|^2$, so the sign on $K_{\rm c}$ is determined by the $\Re(\sigma)$ sign. Since $\sigma$ enters also the second reduced kernel the sign of $K_{\rm t}$ will be determined by the phase difference between pressure perturbation and the rate of change of entropy perturbation. We have therefore computed the phases: $\arg(p\overline{\sigma\,p})$, $\arg(p\overline{\sigma\,s})$ and also $\arg(p\overline{s})$ i.e. the phase between pressure and entropy perturbation. These quantities are shown in \mFig{Fig9} for the same strange mode $S^{-}_4$. From \mFig{Fig9} it is seen that the phase difference between $p$ and $s$ is larger near the maximum opacity bump, but remains there relatively small, and it decreases in H/He~I ionization regions below H ionization zone. This leads to the phases of the sesquilinear products of perturbations entering both reduced kernels to be similar. This in turn leads to the effect that the signs of $K_{\rm c}$ and $K_{\rm t}$ are determined by the signs of the weighting factors $\texttt{A}_4$ and $\texttt{A}_7$. 

To determine whether the near alignment of pressure and entropy perturbation is specific to the examined mode, we computed spectra and obtained phase relations for strange modes in a series of models with $L=10^4\,L_{\odot}$,$M=0.69\,M_{\odot}$ in the effective temperature range $3.7\le \log(T_{\rm eff})\le 3.9$. We found that, for strange modes obtained using the \((3,4)\) form of the outer boundary condition, the pressure perturbation ($p$) becomes in phase with entropy perturbation ($s$) in the H/He~I ionization region for all strange modes found, whether excited or damped. Thus, this phase alignment appears to be a property of the \((3,4)\)-type strange modes. In this case $p\overline{s}$ is approximately real and positive in the H/He~I ionization zone. Therefore
\[
\Re(p\overline{\sigma\,s})\simeq |p|\,|s|\,\Re(\sigma), \qquad \Re(p\overline{\sigma\,p})=|p|^2\Re(\sigma),
\]
and hence
\begin{equation}
\begin{aligned}
	\sgn(K_{\rm c}) &=\sgn(\texttt{A}_4)\,\sgn(\Re(\sigma))=\sgn(\Re(\sigma)),\\
	\sgn(K_{\rm t}) &=\sgn(\texttt{A}_7)\,\sgn(\Re(\sigma))=-\sgn(\Re(\sigma)).
\end{aligned}
\end{equation}
Thus it follows that for strange modes, of the \((3,4)\) type considered here, due to the alignment of pressure and entropy perturbations in the H/He~I ionization region for this type of modes
\begin{equation}
	\sgn(K_{\rm_c})=-\sgn(K_{\rm t}),
\end{equation}
which gives $\sgn(\WC)=-\sgn(\WT)$ and leads to a large cancellation of the contributions to work integral from the two terms $\WC$ and $\WT$. This explains the behavior of $\WC$, $\WT$ and $\WS$ shown in the right panel in \mFig{Fig5}. 

It should be emphasized that \(\WC\) and \(\WT\) are not independent work mechanisms. They are the two contributions obtained when the single pressure volume-rate kernel is expressed through the variables \(p\) and \(s\), using \(q=-A_4p-A_7s\). The cancellation \(\WC\simeq-\WT\) therefore means that, for strange modes, the complete pressure volume-rate work integral \(\WS\) is much smaller than either of its two component contributions. In particular, the pressure--entropy term \(\WT\), which is the closest analogue of the classical work integral, does not by itself represent the net work balance of these modes.

The preceding discussion explains why the work integral \(\WS\) of strange modes is small: it is the residual of two larger projected
contributions. The other side of the balance relation involves the generalized norm \(\NS\), whose behavior differs just as strongly between ordinary and strange modes.

To explain the different behavior of $\NS$ for ordinary and strange modes shown in \mFig{Fig4}, two main factors need be considered - viz. the maximum amplitude of displacement perturbation within the envelope, and the region over which it is localized. 

For the ordinary mode (6-ov) shown in \mFig{Fig6} the displacement perturbation ($|d|$) is localized in a region that extends from the Z-bump, deep in the envelope, to the H/He~I ionization zone, while for the strange mode $S^{-}_4$ the displacement is localized primarily in the H/He~I ionization region with secondary amplitude maximum in the region between the Z-bump and He~II ionization region. There is a substantial difference in amplitude for ordinary and strange modes (see Glatzel (1994), Saio et al. (1998), Saio (2009)), which, for the 15-ov ordinary mode (of frequency similar to that of the $S^{-}_4$ strange mode), reaches values $|d|_{\rm max}\sim 10^3$, while for the strange mode $S^{-}_4$ it is $|d|_{\rm max}\sim 10$. The difference is clearly seen when a plot of $\NC'$ is compared between ordinary and strange modes of similar frequency.

\begin{figure}[htb]
	\includegraphics{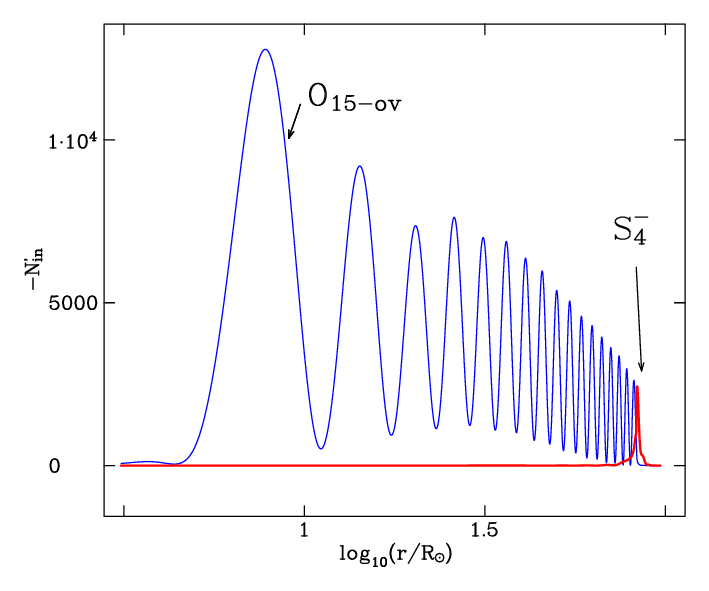}
	\FigCap{The plot of $-\NC'$ for ordinary 15-ov mode and for strange $S^{-}_4$ mode. The displacement perturbation of the strange mode ($|d|$) is much smaller and localized in a much narrower range of radii centered on the H/He~I ionization zones while the amplitude of the ordinary mode is large in a much broader region. This leads to much smaller values for $|\NC'|$ for the strange mode as compared to the ordinary one.}
	\label{fig:Fig10}
\end{figure} 

In \mFig{Fig10} the quantity $-\NC'$ is shown for the 15-ov ordinary and $S^{-}_4$ strange modes. The contribution to the $\NC$ integral is much larger for the ordinary mode. This is due both to its broader localization region, $0.6<x_{10}<1.9$, compared with $1.87<x_{10}<1.95$ for the strange mode in this model, and to the much larger amplitude of $|d|$, which is two orders of magnitude higher for the ordinary mode. 

To explain why $\NS\approx\NT$ for ordinary modes, while $\NS\approx\NC$ for strange modes (\mFig{Fig4}), it is necessary to consider how the localization region of the mode affects the relative sizes of $\NC$ and $\NT$. From \mEq{Ns} it follows that the integrands of both terms contain the common factor $C(x)\texttt{A}_3|d|^2$. Their local ratio may therefore be estimated as
\[
 a=\left|\frac{\NC'}{\NT'}\right|\approx {\texttt{A}_2|\sigma|^2\over 4}.
\]
Since also $\texttt{A}_2\sim r^3$, this gives $a\sim r^3|\sigma|^2$. The inertial contribution is therefore weighted more towards the surface compared to the structural one.

This explains the different behavior of $\NS$ for the two classes of modes. For strange modes, which are localized close to the surface, one expects $|\NC|\gg|\NT|$ and hence $\NS\approx\NC$. For an ordinary mode, whose amplitude is distributed over a much broader, deeper part of the envelope, the structural term dominates and $\NS\approx\NT$. 

Since for strange modes $|\NC|$ is much smaller than for ordinary modes it also follows that $|\NS|$ for strange modes is much smaller than for ordinary ones, as seen in the left panel of \mFig{Fig3}.

\section{Conclusions}

We have formulated the quadratic balance relation of Glatzel (1994) in terms of radial nonadiabatic equations, retaining the boundary contributions. The derivation does not assume weak nonadiabaticity, nor does it rely on the classical cycle-averaged kinetic energy and dissipation formulation. In this paper we have applied our approach to radial pulsations; possible extensions to nonradial modes are left for future work.

We have obtained an identity of the form (see \mEq{BalanceEq})
\[
\WS+\PHS-\PHB=\Re(\sigma)\NS,
\]
where $\WS$ is the work integral (see \mEq{Ws}), $\NS$ is the generalized norm (see \mEq{Ns}) and $\Phi$ are the surface terms given by \mEq{Phi}. This relation is exact within the adopted radial nonadiabatic formulation. 

The balance relation derived here introduces terms absent from the usual near-adiabatic work and energy interpretation
\begin{align*}
\WS&=\WC+\WT, \\
\NS&=\NC+\NT.
\end{align*}
Here $\WT$ is the term closest to the classical pressure-entropy dissipation integral, while $\WC$ is a compressional contribution proportional to $\Re(\sigma)|p|^2$. This term vanishes for strictly adiabatic pulsation $\Re(\sigma)=0$, but plays an important role for pulsations in AGB envelopes.

Similarly, $\NC$ is an inertia-like term, similar to kinetic energy in classical formulation, whereas $\NT$ is a structural-acoustic contribution with no direct counterpart in the classical kinetic-energy integral. It may be noted that the term $\NT$ does not vanish in the adiabatic limit, and in fact turns out to be the dominant term for $\NS$ for ordinary modes in AGB envelopes.

The surface terms $\Phi$ are not optional corrections added to classical expression. They naturally arise in the derivation and are needed for closure of the balance relation. In the present application they also provide a useful diagnostic of the boundary conditions. The term at the lower boundary $\PHB$ is generally small because the inner boundary is being placed in regions where the perturbations decay in strongly nonadiabatic envelopes considered here. The outer boundary term can become large for outer boundary conditions that introduce large perturbations in pulsational quantities at the boundary. For boundary conditions, that do not artificially enhance perturbations near either boundary, both surface terms should be small. We therefore use $\Phi$ as a diagnostic of the adopted form of inner and outer boundary conditions.

Previous studies have emphasized strong nonadiabatic effects (Zalewski (1992), Glatzel (1994), Saio et al. (1998) and Aikawa et al. (1996)), short thermal time scales or radiation pressure effects in strange modes. The present paper does not attempt to establish the nature of strange mode instability i.e. whether strange mode instability should be interpreted as a modified $\kappa$-mechanism, a non-$\kappa$-mechanism, or a local radiation hydrodynamic mode coupling phenomenon. Instead, our approach provides a balance relation diagnostic that any such interpretation should reproduce. 

Using the formulation in this paper we have found that  the dominant role, in the balance equation, may be played not by the classically recognized terms but by the new terms $\NT$ and $\WC$, even for ordinary modes, and particularly in the highly nonadiabatic AGB envelopes. For ordinary modes the generalized norm is mainly determined by the structural term $\NT$,
\[
\NS \approx \NT > 0,
\]
while the work integral is mainly determined by the compressional term,
\[
\WS\approx\WC.
\]
Thus, using our formulation, we have established that ordinary modes in extended envelopes are not described by the classical pair of thermal work plus kinetic energy, and may be properly described only by using all the terms.

The strange modes behave differently from ordinary modes regarding the work and norm integrals. The generalized norm is much smaller in magnitude than for ordinary modes and is given by
\[
\NS\approx\NC<0,
\]
so for strange modes the norm integral is determined by the inertia component. The work integral is also much smaller in magnitude than for ordinary modes, but the work integral is not dominated by a single contribution. For strange modes the two work integral components $\WC$ and $\WT$ are individually large, and of opposite sign:
\[
|\WC|\sim|\WT|,\qquad \sgn(\WC)=-\sgn(\WT).
\]
The total work integral 
\[
\WS=\WC+\WT
\]
is therefore a small residual. This effect is due to a phase relation between pressure ($p$) and entropy ($s$) perturbation that occurs for strange modes in the H/He~I ionization zones: $\arg(p\overline{s})\approx 0$. Due to this effect, found in our analysis and evident for the symmetric strange modes obtained with the \((3,4)\)-type outer boundary conditions, the $\WC$ and $\WT$ largely cancel out. We have found that the signs of these terms follow the sign of $\Re(\sigma)$ according to
\[
\sgn(\WC)=\sgn(\Re(\sigma)),\quad 
\sgn(\WT)=-\sgn(\Re(\sigma)).
\]

It is the near alignment of pressure and entropy perturbations in the H/He~I region that leads to this behavior of the work terms in the balance equation for strange modes and distinguishes their work balance from that of ordinary modes. Within the family of (3,4)-type strange modes examined here, this phase alignment provides a characteristic diagnostic signature of their work balance; it does not by itself establish their physical origin.

It is also worth noting that besides the ordinary and strange modes discussed in this paper the balance relation holds for thermal modes and can be used in their analysis as it does not invoke any notion of cycle-averaging.

The balance relation derived here provides both the check on the real part of the computed eigenfrequency, \(\Re(\sigma)\), and a diagnostic of the terms through which different classes of modes satisfy the global work-norm balance. It shows that classical dissipation-kinetic energy formulation omits terms that are not only important for strange and thermal modes, but which can also determine the balance of ordinary modes in extended envelopes.

A nonradial extension of the balance relation, including the additional horizontal and gravitational-potential coupling terms, is currently in preparation and will be presented separately.

\end{document}